\documentclass{article}
\usepackage{filecontents}
\usepackage{url}
\usepackage[utf8]{inputenc}
\usepackage{graphicx}
\usepackage[normalem]{ulem}
\usepackage[table]{xcolor}
\usepackage{braket}
\usepackage[margin=1in]{geometry}
\usepackage{color}
\usepackage{xintexpr}
\usepackage{jheppub}
\usepackage{siunitx}
\usepackage[noabbrev]{cleveref}

\preprint{
\begin{flushright}
TTK-22-13\\ TTP22-018 \\ P3H-22-032 \\ FERMILAB-PUB-22-238-T
\end{flushright}}
\arxivnumber{}

\title{Searching for dark radiation at the LHC}

\author[1]{Elias Bernreuther,}
\author[2,3]{Felix Kahlhoefer,}
\author[2]{Michele Lucente}
\author[2]{and Alessandro Morandini}

\affiliation[1]{Fermi National Accelerator Laboratory, Batavia, IL 60510, USA}
\affiliation[2]{Institute for Theoretical Particle Physics and Cosmology (TTK), RWTH Aachen University, D-52056 Aachen, Germany}
\affiliation[3]{Institut f\"{u}r Theoretische Teilchenphysik, Karlsruhe Institute of Technology (KIT), 76128 Karlsruhe, Germany}

\emailAdd{ebernreu@fnal.gov}
\emailAdd{kahlhoefer@physik.rwth-aachen.de}
\emailAdd{lucente@physik.rwth-aachen.de}
\emailAdd{morandini@physik.rwth-aachen.de}

\abstract{In this work we explore the intriguing connections between searches for long-lived particles (LLPs) at the LHC and early universe cosmology. We study the non-thermal production of ultra-relativistic particles (i.e. dark radiation) in the early universe via the decay of weak-scale LLPs and show that the cosmologically interesting range $\Delta N_\text{eff} \sim 0.01\text{--}0.1$ corresponds to LLP decay lengths in the mm to cm range. These decay lengths lie at the boundary between prompt and displaced signatures at the LHC and can be comprehensively explored by combining searches for both. To illustrate this point, we consider a scenario where the LLP decays into a charged lepton and a (nearly) massless invisible particle. By reinterpreting searches for promptly decaying sleptons and for displaced leptons at both ATLAS and CMS we can then directly compare LHC exclusions with cosmological observables. We find that the CMB-S4 target value of $\Delta N_\text{eff}=0.06$ is already excluded by current LHC searches and even smaller values can be probed for LLP masses at the electroweak scale.}


\begin{document}

\maketitle

\section{Introduction}

Recent years have seen a rapid growth of interest in long-lived particles (LLPs) with a mass around the electroweak scale and a proper decay length between millimetres and metres, which could give rise to a wide range of exciting signatures at the Large Hadron Collider (LHC)~\cite{Alimena:2019zri}. The interest in these particles stems not only from their unusual experimental signatures but also from an intriguing connection to early universe cosmology: A proper decay length of $1\,\mathrm{cm}$ corresponds to a decay width of around $10^{-14}\,\mathrm{GeV}$, which is comparable to the Hubble expansion rate of the universe (in standard cosmology) at temperatures around $100\,\mathrm{GeV}$. The implication is that whatever particles are produced in the decays of the LLPs at the LHC would also have been produced efficiently in the early universe and would have affected its subsequent evolution~\cite{Kahlhoefer:2018xxo}.

Particular attention has been paid to the case that the LLP decays involve massive stable particles with negligible couplings to SM states. These particles would obtain a sizeable abundance in the early universe via the so-called freeze-in mechanism~\cite{Hall:2009bx,Bernal:2017kxu} and may account for the observed dark matter density~\cite{Calibbi:2018fqf,Belanger:2018sti}. A closer inspection, however, reveals that the decay width of the LLP required to reproduce observations must be significantly smaller than the Hubble rate at the electroweak scale, such that the corresponding decay lengths are large compared to typical LHC scales~\cite{No:2019gvl}. Various proposals have explored possible modifications of this argument, for example if the DM mass is at the keV scale~\cite{DEramo:2020gpr,Decant:2021mhj,Li:2021okx} or if the universe undergoes an early period of matter domination~\cite{Cosme:2020mck,Calibbi:2021fld}. The conclusion is that, while the observed dark matter abundance may be reproduced, it is hardly possible to obtain concrete predictions for LLP searches at the LHC from this argument alone.

In the present work we instead explore the possibility that the LLP decays involve massless particles, which would act in the early universe as dark radiation (DR). Since the energy density of DR decreases more rapidly with decreasing temperature than the one of DM, the former may significantly exceed the latter in the very early universe. The corresponding LLP decay rates may therefore be significantly larger than for the case of DM production. Indeed, it turns out that current cosmological bounds on the number of additional relativistic degrees of freedom, $\Delta N_\text{eff} < 0.2$ at 95\% confidence level~\cite{Planck:2018vyg, Fields:2019pfx}, place virtually no constraints on this scenario, in the sense that even a fully thermalised species (and hence an arbitrarily large decay width) is allowed as long as the LLP decays happen sufficiently early that the energy density of DR gets diluted before recombination. 

The next generation of missions to study the Cosmic Microwave Background (CMB) may however change this picture decisively. Indeed, the sensitivity of CMB-S4~\cite{CMB-S4:2016ple,Abazajian:2019eic} is expected to be sufficient to exclude the case of thermalised DR and may even provide hints for non-thermal DR~\cite{Fields:2019pfx}. At the same time, (self-interacting) DR is an important ingredient for various models that attempt to resolve the so-called Hubble tension, i.e.\ the discrepancy between various early-time and late-time measurements of the Hubble constant~\cite{Blinov:2020hmc,Aloni:2021eaq}.

In this work we study in detail the freeze-in production of DR in the early universe~\cite{Hasenkamp:2012ii}, including relativistic and quantum corrections (studied previously for the case of freeze-in production of dark matter in Refs.~\cite{Belanger:2018ccd,Lebedev:2019ton,Biondini:2020ric,Bringmann:2021sth}) and the backreaction from inverse decays. We show that this scenario is highly predictive and that the most interesting regions of parameter space ($\Delta N_\text{eff} \sim 0.01\text{--}0.1$) correspond to LLPs with a decay length of the order of $1\text{--}10\,\mathrm{mm}$. These decay lengths lie at the often overlooked boundary between searches for promptly decaying particles and searches for LLPs~\cite{Ito:2017dpm}. A key part of our study is therefore to understand how the sensitivity of prompt searches changes for non-negligible decay lengths and to explore the complementarity of searches for prompt and displaced decays. Indeed, we show the combination of these searches  possesses sufficient sensitivity for LLPs to achieve a complete coverage of the interesting range of decay lengths.

While the general mechanism discussed in this work applies to a wide range of models, we focus for concreteness on the case of a scalar LLP with electroweak charges that decays into a charged lepton and an invisible DR particle. The most relevant prompt searches are therefore those for the direct production of sleptons~\cite{ATLAS:2019lff,CMS:2020bfa}, whereas LLPs with sizeable decay lengths are constrained by searches for displaced leptons~\cite{ATLAS:2020wjh,CMS:2021kdm}. We perform a detailed reinterpretation of these searches in order to identify the allowed regions of parameter space. Future constraints on $\Delta N_\text{eff}$ (or hints of a non-zero value) can then be used to further bound this parameter space and relate the mass and lifetime of the LLP.

The remainder of this work is structured as follows. In section~\ref{sec:freeze-in} we introduce the model that we consider and derive the contribution to $\Delta N_\text{eff}$ from the freeze-in mechanism. We then discuss the reinterpretation of LHC searches in the context of our model in section~\ref{sec:LHC} and present the resulting constraints. In section~\ref{sec:conclusions} we then combine these two approaches in order to compare the constraints from cosmology and the LHC and conclude.

\section{Freeze-in production of dark radiation}
\label{sec:freeze-in}

Although we will keep the discussion in this section as general as possible, it will be helpful to introduce a specific model for concreteness. Let us therefore consider a scalar boson $B = (B_e, B_\mu, B_\tau)^\mathrm{T}$ with three different flavour states of equal mass $m_B$ and hypercharge $Y_B = -1$ as well as a Majorana fermion $\chi$ that is a singlet under the Standard Model (SM) gauge group. Both particles are assumed to be odd under a $\mathbb{Z}_2$ symmetry, such that the only allowed renormalisable interaction of $\chi$ is given by
\begin{equation}
\mathcal{L}_\text{int} = B^\mathrm{T} \cdot y_\ell \cdot (\bar{\ell}_\mathrm{R}\, \chi)\, +\textrm{h.c.}\;,
\end{equation} 
where $\ell_R = (e_R, \mu_R, \tau_R)^\mathrm{T}$ denotes the right-handed SM leptons and $y_\ell = \text{diag}(y_e, y_\mu, y_\tau)$ is the coupling matrix. 
This interaction corresponds to the one between right-handed sleptons and neutralinos in supersymmetric extensions of the SM. Unless explicitly stated otherwise, we assume flavour-universal couplings, i.e.\ $y_e = y_\mu = y_\tau \equiv y$, in the following.  The scalar boson $B$ has additional gauge interactions, which keep it in equilibrium with the SM thermal bath in the early universe and allow for sizeable production rates at the LHC. The only way to produce $\chi$ particles, on the other hand, is through the decays of $B$. The $\mathbb{Z}_2$ symmetry then ensures that $\chi$ is stable and can act as DR in the early universe. 

In the following we will assume that the mass of $\chi$ is negligible during recombination, which implies $m_\chi \ll 1 \, \mathrm{eV}$. 
However, for $m_\chi \gtrsim 1 \, \mathrm{meV}$, these particles would be non-relativistic in the present universe and therefore affect structure formation in a similar way as hot dark matter. It has been shown in Ref.~\cite{Baur:2017stq}, that for masses below the keV scale constraints from structure formation become independent of the specific mass value, and result into an upper bound on the fraction $F_\text{HDM}$ of hot dark matter relative to the total dark matter abundance. The exact upper bound on $F_\text{HDM}$ depends on the production mechanism of the hot DM component (which in turns affects the DM free streaming length), but one can safely assume that the bounds are not relevant as long as $F_\text{HDM}$ is much smaller than the percent level, which is always the case for $m_\chi < 0.1 \, \mathrm{eV}$ and $\Delta N_\text{eff} < 0.1$.

If the coupling $y$ is sufficiently large, and as long as $T \gtrsim m_B$, the DR energy density $\rho_\chi$ will follow an equilibrium distribution:
\begin{equation}
 \rho_\chi = g_\chi \frac{7}{8} \frac{\pi^2}{30} T^4 \; ,
\end{equation}
where $g_\chi = 2$ denotes the degrees of freedom of $\chi$.
Once the temperature becomes much smaller than $m_B$, the DR will decouple from the SM thermal bath and evolve independently. This means that its energy density will simply redshift, such that
\begin{equation}
 Z_\chi \equiv \frac{\rho_\chi(x)}{s^{4/3}(x)} = \text{const} \; ,
\end{equation}
where $s(x)$ denotes the entropy density of the SM thermal bath and $x = m_B / T$.

As heavy particles in the plasma become Boltzmann suppressed and annihilate away, they transfer their entropy to lighter species in the plasma but not to DR. As a result we find that for fully decoupled DR~\cite{Blennow:2012de}
\begin{equation}
 \frac{\rho_\chi}{\rho_\gamma} \propto g_s^\ast(x)^{4/3}
\end{equation}
where $\rho_\gamma$ denotes the energy density of photons and $g^\ast_s(x)$ denotes the number of entropy degrees of freedom. We can therefore express the contribution of $\chi$ to the effective number of relativistic degrees of freedom as\footnote{We note that there is no unique way to define $\Delta N_\text{eff}$ before neutrino decoupling. The definition adopted here can be interpreted as the projected value of $\Delta N_\text{eff}$ in the present universe under the assumption that DR is neither produced nor destroyed between $x$ and today, implying $\rho_\chi^0=\rho_\chi(x) (s^0/s(x))^{4/3}$. After neutrino decoupling, corresponding to the temperature range probed by observations, $g_s^\ast(x) = g^\ast_{s,0}$ and hence we recover the standard definition of $\Delta N_\text{eff}$ as the ratio of the energy density in dark radiation and the energy density of a single neutrino species in the instant-decoupling approximation.}:
\begin{equation}\label{eq:dneff}
\Delta N_\text{eff}(x) = \frac{\rho_\chi(x)}{\frac{7}{8} \left(\tfrac{4}{11}\right)^{4/3} \rho_{\gamma}(x)} \left(\frac{g_{s,0}^\ast}{g_s^\ast(x)}\right)^{4/3} = \frac{Z_\chi(x) \, s_0^{4/3}}{\frac{7}{8} \left(\tfrac{4}{11}\right)^{4/3} \rho_{\gamma,0}} \; ,
\end{equation}
where the subscript 0 denotes present-day quantities, i.e.\ $g^\ast_{s,0} = 3.9$. If $B$ is sufficiently heavy compared to all SM particles, we can approximately take $g^\ast_s(x) \approx 100$ at the time when DR decouples, leading to
\begin{equation}
 \Delta N_\text{eff}(x) \approx 0.05 \; ,
\end{equation}
which is well below the current bound $\Delta N_\text{eff} < 0.2$ from a combination of data from the CMB and BBN. In other words, as long as DR decouples from the SM thermal bath before the QCD phase transition there are no cosmological constraints on the coupling $y$.

With future cosmological observations it may, however, be possible to probe values of $\Delta N_\text{eff}$ as small as 0.05 and therefore potentially exclude any form of DR that enters into equilibrium with the SM thermal bath at some point in the cosmological history. The implication would then be that $y$ must be small enough to ensure that $\chi$ does not thermalise with the other particles. To first approximation this requirement can be expressed as
\begin{equation}
 \Gamma_B < H(x = 1) \; , \label{eq:naive}
\end{equation}
where
\begin{equation}
\Gamma_B = \frac{|\mathcal{M}|^2}{16 \pi m_B} = \frac{y^2 m_B}{16\pi}
\end{equation}
is the decay width of the process $B \to \ell + \chi$ (averaged over flavours and assuming $m_B \gg m_\ell$) and $H(x) = 1.66 \sqrt{g^\ast(x)} m_B^2 / (x^2 M_\text{P})$ is the Hubble rate during radiation domination with $g^\ast$ being the number of energy degrees of freedom and $M_\text{P}$ denoting the Planck mass. For $m_B \sim 200 \, \mathrm{GeV}$ this requirement translates to $y \lesssim 10^{-7}$.

In the following we will refine this estimate and at the same time calculate the energy density $\rho_\chi$ also for the case that $\chi$ does not enter into thermal equilibrium with the SM thermal bath. For this purpose, we need to consider the Boltzmann equation describing the evolution of the DR phase space density $f_\chi$:
\begin{equation}
 E \frac{\partial f_\chi}{\partial t} - H p^2 \frac{\partial f_\chi}{\partial E} = \hat{C}[f_\chi] \; ,
\end{equation}
where the general collision operator for (inverse) decays is given by
\begin{align}
 \hat{C}[f_\chi] = \frac{1}{2 g_\chi} \int \frac{\mathrm{d}^3 p_\ell}{(2\pi)^3 2 E_\ell} \int \frac{\mathrm{d}^3 p_B}{(2\pi)^3 2 E_B} & (2\pi)^4 \delta^{(4)}(p_B - p_\ell - p_\chi) \nonumber \\ & \times \left(|\mathcal{M}|^2_{B\to\chi\ell}f_B (1-f_\chi)(1-f_\ell)-|\mathcal{M}|^2_{\chi\ell\to B}f_\chi f_\ell (1+f_B)\right) \; ,
\end{align}
where $p_X$ and $f_X$ denote the four-momentum and phase space density of particle species $X$.

In the context of dark matter relic density calculations it is common to integrate this expression over $\mathrm{d}^3 p_\chi / E_\chi$ in order to obtain a differential equation for the number density $n_\chi$. In the present context we instead calculate the first moment of the Boltzmann equation, i.e.\ we integrate over $\mathrm{d}^3 p_\chi$ to obtain a differential equation for $\rho_\chi$. The left-hand side of the Boltzmann equation then becomes
\begin{equation}
 g_\chi \int \frac{\mathrm{d}^3 p_\chi}{(2\pi)^3} \left(E \frac{\partial f_\chi}{\partial t} - H p^2 \frac{\partial f_\chi}{\partial E} \right) = \dot{\rho}_\chi + 3 H \rho_\chi + H n_\chi \biggl< \frac{p^2}{E} \biggr> = x \tilde{H} s^{4/3} \frac{\mathrm{d}Z_\chi}{\mathrm{d} x}
\end{equation}
with
$\tilde H(x)=H(x)(1-\frac{1}{3}\frac{\mathrm{d}\log g^*(x)}{\mathrm{d} \log x})$ as in Ref.~\cite{degrees}. To simplify the right-hand side, we make use of the fact that both $f_B$ and $f_\ell$ are given by their respective equilibrium distributions\footnote{The case that the parent particle deviates from an equilibrium distribution before decaying has recently been studied in great detail in Ref.~\cite{Decant:2021mhj}.}
\begin{align}
f^\text{eq}_B(E_B, T) = \frac{g_B}{e^{E_B/ T} - 1} \;, \qquad f^\text{eq}_\ell(E_\ell, T) = \frac{g_\ell}{e^{E_\ell/T} + 1}
\end{align}
with $g_B = g_\ell = 6$ (including flavours) and that in thermal equilibrium the rate for $B \to \chi + \ell$ must be equal to the rate of $\chi + \ell \to B$. Carrying out the integration over $\mathrm{d}^3 p_B$ and further integrating over $\mathrm{d}^3 p_\chi$ we then obtain
\begin{align}
 g_\chi \int \frac{\mathrm{d}^3 p_\chi}{(2\pi)^3} \hat{C}[f_\chi] & = \frac{m_B \, \Gamma_B}{8\pi^4} \int \frac{d^3p_\chi}{2E_\chi} \int \frac{d^3p_\ell}{2E_\ell} \frac{2 E_\chi}{E_B} \left(1 - \frac{f_\chi(E_\chi)}{f_\chi^\text{eq}(E_\chi)}\right) f^\text{eq}_B(E_B) \delta(E_B - E_\chi - E_\ell) \nonumber \\
 & \equiv \frac{m_B \, \Gamma_B}{8\pi^4} I\;.
\end{align}

It is important to note that since we have integrated over $\mathrm{d}^3 p_\chi$ rather than $\mathrm{d}^3 p_\chi / E_\chi$, the integral $I$ is not Lorentz-invariant. Moreover, the equilibrium distributions $f_\chi^\text{eq}$ and $f_B^\text{eq}$ take a simple form only in the cosmic rest frame. Nevertheless, we can simplify the Lorentz-invariant expression $\delta(E_B - E_\chi - E_\ell) / (2 E_B)$ by switching to the centre-of-mass (cms) frame, in which the decaying particle is at rest. To do so, we make use of the fact that in a general frame $f_\chi^\text{eq}(E_\chi)$ becomes $f_\chi^\text{eq}(u \cdot k_\chi)$, where $k_\chi = (\omega_\chi, \mathbf{k}_\chi)$ denotes the four-momentum of $\chi$ in that frame and $u$ denotes the four-momentum of the cosmic fluid, which is (1, 0, 0, 0) in the cosmic rest frame. To transform into the cms frame, we follow Ref.~\cite{Arcadi:2019oxh,DeRomeri:2020wng} and define $p = (p_\chi + p_\ell)/2$ and $k = (p_\chi - p_\ell)/2$ and introduce the variables $E, \eta$ and $\theta$ such that in the cosmic rest frame $p^0 = E \cosh \eta$ and $p^3 = E \sinh \eta \cos \theta$.\footnote{To fully determine all variables, we also need $p^1 = E \sinh \eta \sin \theta \cos \phi$ and $p^2 = E \sinh \eta \sin \theta \sin \phi$, but these will play no role here.} Here $\eta$ denotes the rapidity of the cms frame in the cosmic rest frame and $\theta$ denotes the direction of the boost. One can then show that the energy of $\chi$ in the cosmic rest frame $E_\chi = \omega_\chi \cosh \eta + |\mathbf{k}_\chi| \cos \theta \sinh \eta$. Assuming massless decay particles, this becomes
\begin{equation}
 E_\chi = E(\cosh \eta + \sinh \eta \cos \theta) \; .
\end{equation}
After an appropriate transformation of the integration measure, we obtain
\begin{align}
I =  8 \pi^2 \int dE \, E^3 \int d\eta \, \sinh^2 \eta \int d\cos\theta & \frac{2 E(\cosh \eta + \sinh \eta \cos \theta)}{m_B} f_B(2 E \cosh \eta) \delta(E - m_B/2) \nonumber \\
& \times  \left[1 - \frac{f_\chi\left(E(\cosh \eta + \sinh \eta \cos \theta)\right)}{f_\chi^\text{eq}\left(E(\cosh \eta + \sinh \eta \cos \theta)\right)}\right] \; ,
\end{align}
where we have already performed the integration over the remaining angular variables. Performing the integration over $E$ yields
\begin{align}
I = \pi^2 m_B^3 \int d\eta \, \sinh^2 \eta \int d\cos\theta & (\cosh \eta + \sinh \eta \cos \theta)  f_B(m_B \cosh \eta)  \nonumber \\ & \times \left[1 - \frac{f_\chi\left(\tfrac{m_B}{2}(\cosh \eta + \sinh \eta \cos \theta)\right)}{f_\chi^\text{eq}\left(\tfrac{m_B}{2}(\cosh \eta + \sinh \eta \cos \theta)\right)}\right]\; .
\label{eq:I}
\end{align}

Eq.~\eqref{eq:I} can in principle be evaluated for any given phase space distribution $f_\chi(E_\chi)$. Here we will make the simplifying assumption that the DR energy density can be characterised by a temperature $T_\chi$, which may be different from the photon temperature $T$, and that the chemical potential vanishes\footnote{This assumption is sensible provided there are sufficiently strong self-interactions between DR particles that the DR is in kinetic and chemical equilibrium with itself. But even in the absence of such interactions, we expect the assumption to be valid to good approximation for $f_\chi$ close to its equilibrium distribution. For $f_\chi \ll f_\chi^\text{eq}$, on the other hand, the contribution from the inverse process becomes negligible and the precise functional form of $f_\chi$ is irrelevant.}, such that
\begin{equation}
 f_\chi(E_\chi) = \frac{1}{e^{E_\chi / T_\chi} + 1} \; .
\end{equation}
We can then introduce the dimensionless parameters $\epsilon \equiv 1 - T_\chi / T = 1 - (\rho_\chi / \rho_\chi^\text{eq})^{1/4}$ and $z =2 E_\chi / m_B$ in order to define the ratio
\begin{equation}
 r_\chi(z;T,\epsilon) = \frac{f_\chi(\frac{m_B z}{2};T,\epsilon)}{f_\chi^\text{eq}(\frac{m_B z}{2};T)} = \frac{e^{\tfrac{m_B z}{2T}} + 1}{e^{\tfrac{m_B z}{2T (1-\epsilon)}} + 1} \; .
\end{equation}
Introducing $w = \cosh \eta$, the integrated Boltzmann equation then becomes
\begin{align}
\tilde H x s^{4/3}(x) \frac{\mathrm{d} Z_\chi}{\mathrm{d}x}=\frac{m_B^4 \Gamma_B}{8\pi^2}\int_{-1}^1\mathrm{d}\cos\theta\int_1^{\infty}\mathrm{d}w & \sqrt{w^2-1}f_B^{eq}(w;T) (w+\sqrt{w^2-1}\cos\theta) \nonumber \\ & \times \left[1-r_\chi\left(w+\sqrt{w^2-1}\cos\theta;T,\epsilon\right)\right]\; .
\label{eq:Boltzmann}
\end{align}
We find that for $\epsilon \ll 1$ the right-hand side becomes proportional to $\epsilon$, corresponding to the expected behaviour that the Boltzmann equation restores equilibrium.

We note that in the absence of backreaction (setting $r_\chi = 0$) the integral over $\cos \theta$ can be performed analytically, giving
\begin{align}
 I & = 2 \pi^2 m_B^3 \int \mathrm{d} w \sqrt{w^2 - 1} w f_B(m_B w) = \frac{\pi}{2} \int \mathrm{d}^3 \, p_B f_B(E_B) = 4 \pi^4 \, n_B\; ,
\end{align}
where we have introduced $E_B = m_B w$ in the second line. In other words, the relativistic corrections for freeze-in from decays cancel out, and the integrated collision operator is simply given by $m_B \Gamma_B n_B / 2$.

We emphasize that the result in eq.~\eqref{eq:Boltzmann} is very general. In particular, we have not made the usual approximation that $B$ is non-relativistic when it decays~\cite{Heeck:2017xbu} and we have fully accounted for quantum statistics in our calculation. Our result can be generalised to different decay processes by making the obvious sign replacements (if the decaying particle is a fermion or if the DR particle is a boson) and adjusting the degrees of freedom. If several different decay processes contribute to the production of DR, the right-hand side simply becomes the sum over all of these processes.\footnote{Note that, in principle, there may also be a contribution from $2 \to 2$ processes such as $Z + B \to \ell + \chi$. However, if the phase space for the decay $B \to \ell + \chi$ is unsuppressed, these processes are found to give a negligible contribution to the freeze-in production~\cite{Biondini:2020ric}.}

\begin{figure}
\centering
\includegraphics[width=\textwidth]{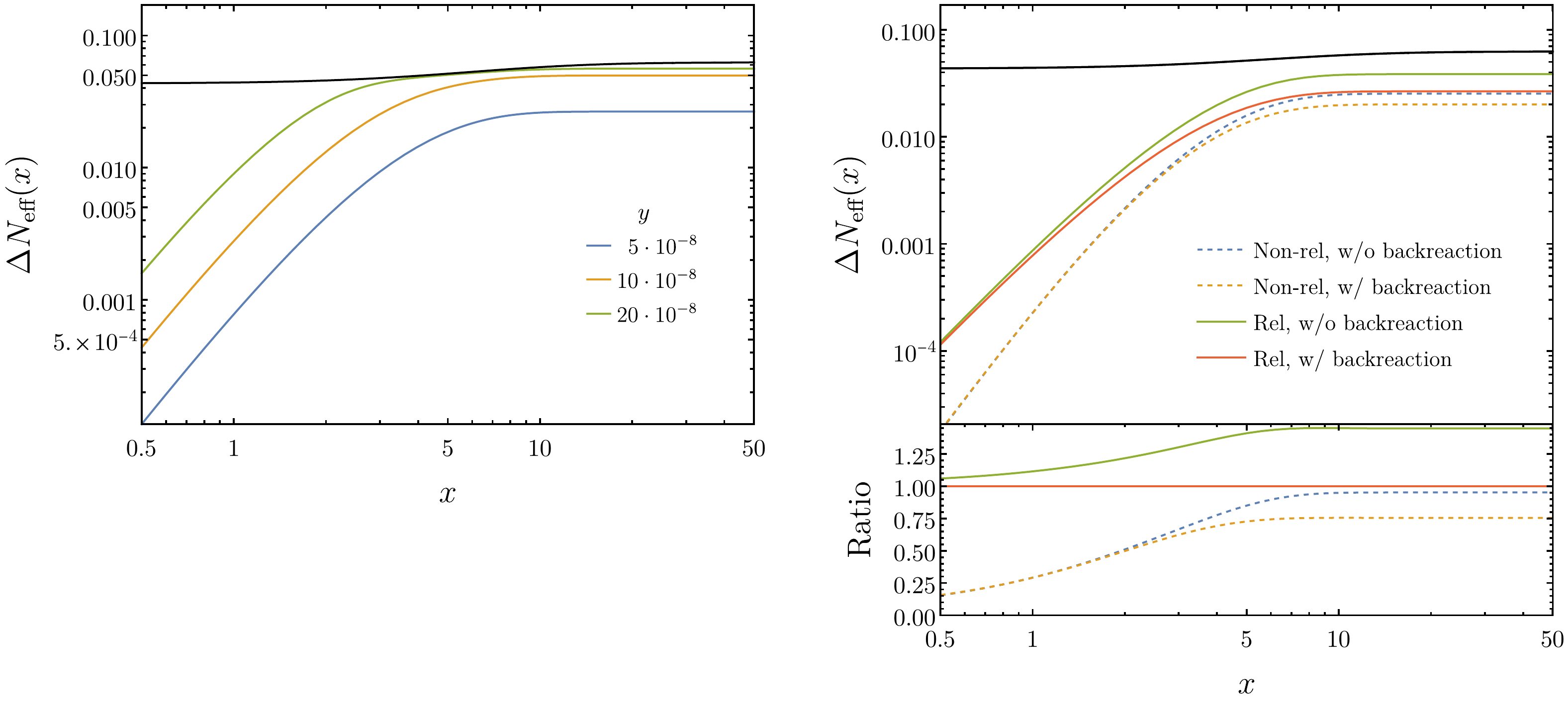}
\caption{Shift in the effective number of relativistic species as defined in eq.~\eqref{eq:dneff} as a function of time for a benchmark mass value of $m_B=\SI{300}{\giga\electronvolt}$. The left panel shows how increasing the coupling increases $\Delta N_\text{eff}$, but the increase becomes milder for large couplings due to the backreaction effect. The right panel illustrates the importance of the various effects included in our analysis compared to simpler approximations (non-relativistic decays, no backreaction).}\label{fig:eq300}
\end{figure} 

The integral in eq.~\eqref{eq:Boltzmann} can be evaluated numerically\footnote{We provide the tabulated integrals for different scenarios as supplementary files.} in order to solve the Boltzmann equation under the assumption that the initial energy density of DR is negligible. Once the solution of the Boltzmann equation is obtained, the value of $\Delta N_\text{eff}$ follows directly from eq.~\eqref{eq:dneff}. In the left panel of \cref{fig:eq300} we show the evolution of $\Delta N_\text{eff}$ for $m_B=\SI{300}{\giga\electronvolt}$ for different values of $y$. For small couplings $y \lesssim 5 \cdot 10^{-8}$ we find that $Z_\chi$ grows with $x$ until about $x \approx 5$, when the parent particles become Boltzmann suppressed and $Z_\chi$ approaches a constant value. In this regime the final energy yield is proportional to $y^2$. For larger values of $y$, on the other hand, $Z_\chi$ approaches the equilibrium value (indicated by the black line) and a further increase in $y$ does not imply a correspondingly larger energy yield. We note that this saturation happens for slightly smaller values of $y$ than suggested by the naive estimate in eq.~\eqref{eq:naive}. However, even for large couplings $Z_\chi$ only traces the equilibrium energy density as long as the parent particles are abundant in the plasma. At low temperatures $Z_\chi$ always approaches a constant, while the equilibrium value increases slightly as the number of entropy degrees of freedom in the plasma decreases.

To illustrate the importance of the various effects that we have included in our calculation, we show in the right panel of \cref{fig:eq300} the curves that would be obtained when neglecting backreaction and when not taking into account the statistical properties of the various particles. We emphasize that the role of backreactions is considerable even if the comoving abundance always remains well below the equilibrium value. As expected, including the backreaction effect always reduces the final yield. In contrast, the role of the relativistic corrections is more subtle and ultimately depends not only on the spin nature of the particles involved but also on the mass and couplings taken into consideration. When considering the decay as relativistic, both the bosonic nature of the parent particle and the fermionic nature of the daughter particle are important. These two effects play opposite roles and are relevant at different temperatures as can be understood by looking at the lower panel on the right-hand side of \cref{fig:eq300}. At higher temperatures the bosonic nature of the parent particle plays the most important role and it increases the abundance $Z_\chi$ with respect to the case where the distribution of $B$ is approximated by a Maxwell-Boltzmann distribution. The fermionic suppression due to the  statistics of the daughter particle is relevant if the backreactions play a role and is hence relevant at lower temperatures, provided the abundance of $\chi$ is high enough.

\begin{figure}
\centering
\includegraphics[width=0.49\textwidth]{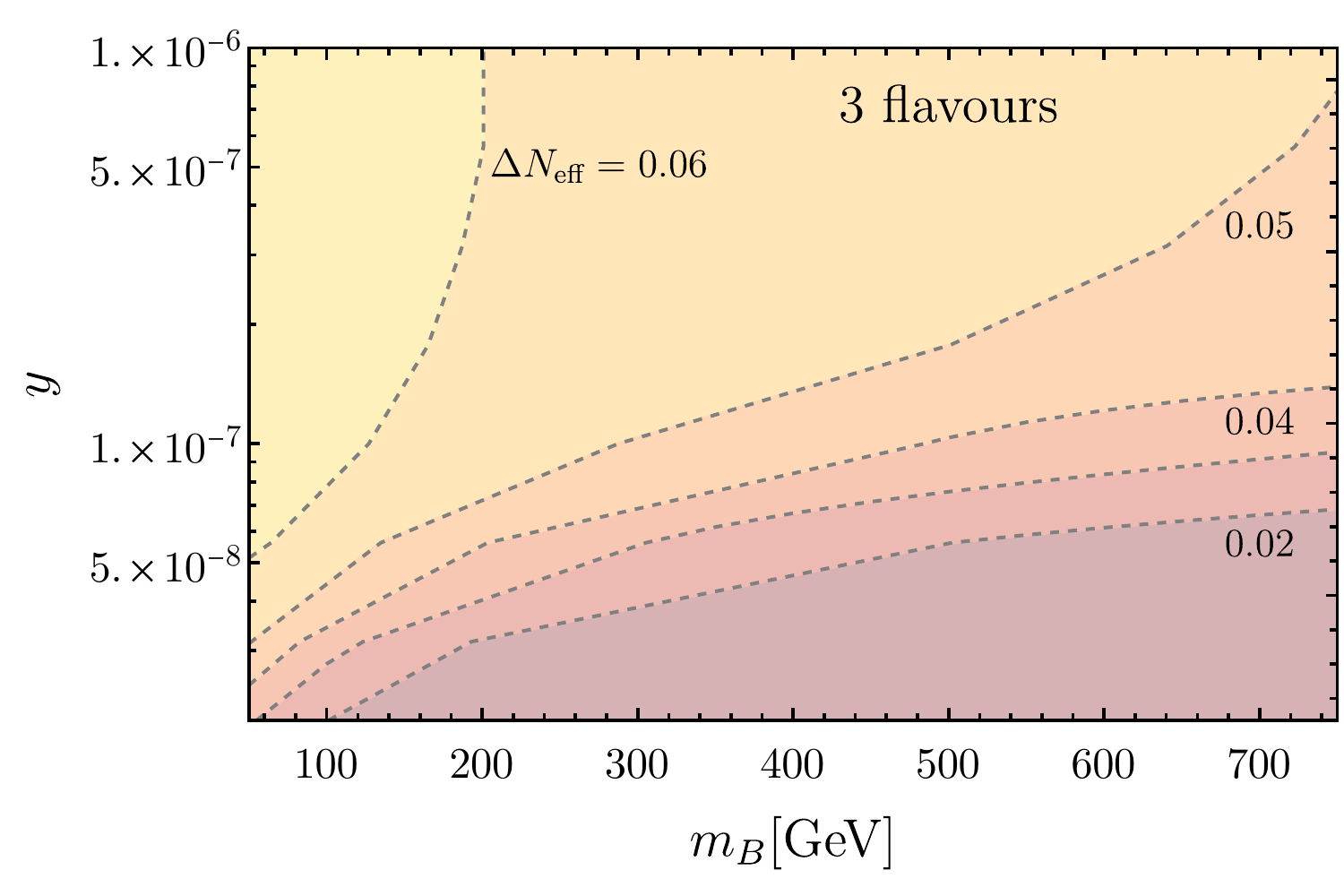}
\includegraphics[width=0.49\textwidth]{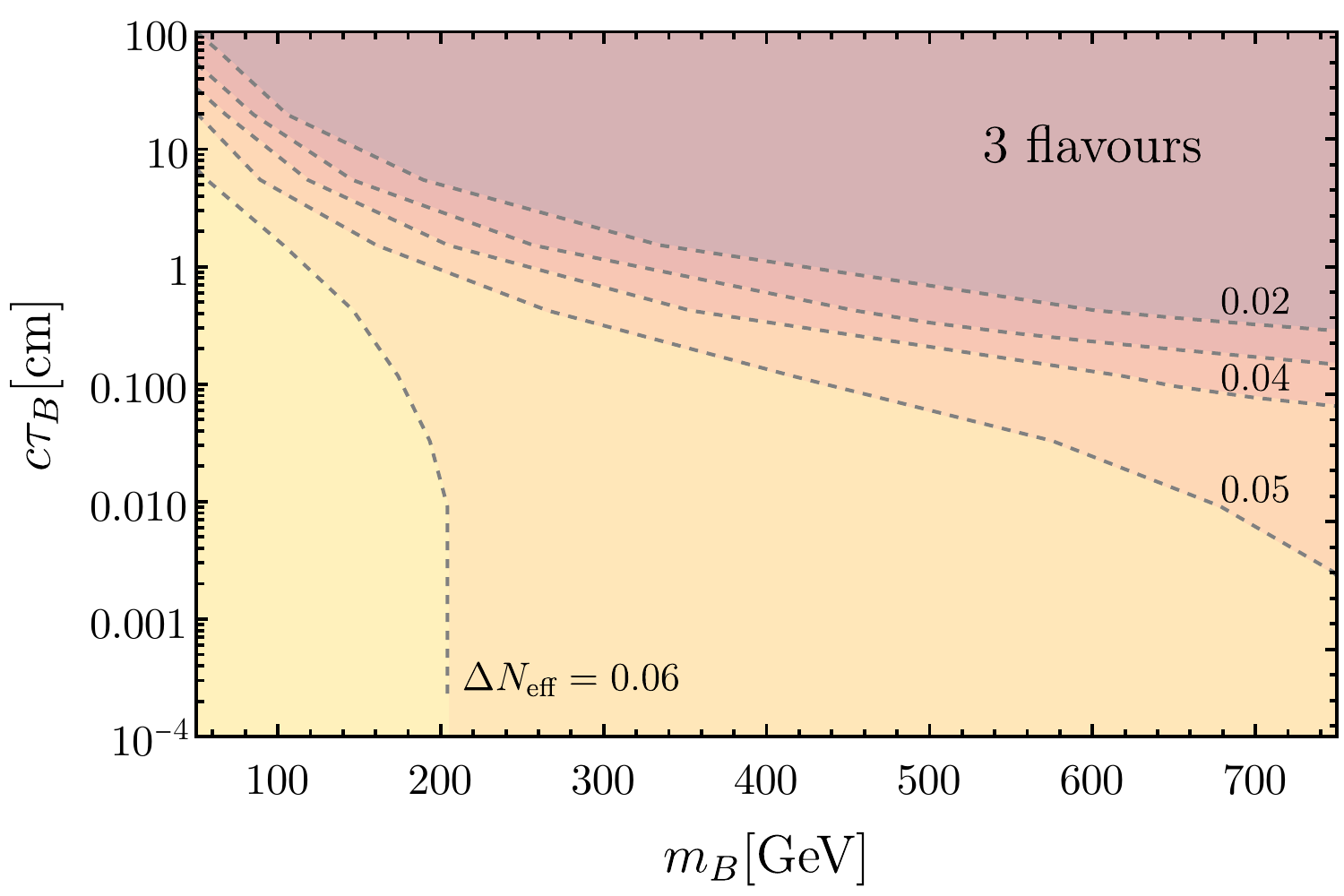}
\caption{Shift in the effective degrees of freedom as a function of the mass and coupling (\textit{left}) or proper decay length (\textit{right}). Shorter lifetimes correspond to larger coupling and for short enough lifetimes we reach the equilibrium densities and the upper bound on $\Delta N_\text{eff}$.}\label{fig:lepscan}
\end{figure}

Since the model that we consider is characterised by only two parameters (the mass $m_\chi$ being negligible), we can easily scan over the parameter space and calculate the resulting contribution to $\Delta N_\text{eff}$. The results are shown in \cref{fig:lepscan} as a function of $m_B$ and either the coupling $y$ (left panel) or the LLP decay length (right panel). In the parameter regions investigated we find $\Delta N_\text{eff}\leq 0.07$, with higher values for smaller masses and larger couplings (shorter lifetimes). For large enough couplings we see a saturation due to strong backreaction such that one cannot exceed the equilibrium values. We also observe that the corresponding decay lengths are macroscopic, in the sense that they are of the order of $1\text{--}\SI{10}{\milli\meter}$. 

\section{LHC signatures}\label{sec:LHC}

In the previous section we were able to
map the LLP decay length to predictions for $\Delta N_{\text{eff}}$ that may be observable with the next generation of CMB experiments. Now that we know the parameter space of interest for DR, we want to see which LHC searches are able to probe it. By examining \cref{fig:lepscan} we can see that we cannot rely exclusively on either prompt signatures or LLP signatures but we will have to consider both. To understand which searches are the most constraining for our model, we will have to carefully consider how the sensitivity of prompt signatures is modified in the case that the lepton tracks are not produced at the interaction vertex but with a macroscopic impact parameter. 

To first approximation, we can assume that a prompt search remains valid if the LLP decay happens within some small distance $\Delta x$ from the interaction point. For an LLP with velocity $\beta = v /c$, boost factor $\gamma$ and proper decay length $c \tau$, the probability to decay within this distance is given by
\begin{equation}
 p(x < \Delta x) = \frac{1}{\beta \gamma c \tau} \int_0^{\Delta x} \exp \left( - \frac{x}{\beta \gamma c \tau} \right) \mathrm{d}x = 1 - \exp \left( - \frac{\Delta x}{\beta \gamma c \tau} \right) \approx \frac{\Delta x}{\beta \gamma c \tau} \; ,
\end{equation}
where the final step is valid for $\Delta x \ll \beta \gamma c \tau$. We therefore expect the number of predicted events in prompt searches to decrease proportionally to $(c \tau)^{-1}$ for large proper decay lengths. The constant of proportionality depends on the kinematic distribution of the LLPs, which can only be extracted from Monte Carlo simulations.

As input for our simulations we use a UFO file for our model created with \textsc{FeynRules}~\cite{Alloul:2013bka}. For both prompt and LLP signatures we generate events with \textsc{MadGraph5\_aMC{@}NLO}~2.8.2~\cite{Alwall:2014hca} interfaced with \textsc{Pythia~8}~\cite{Sjostrand:2014zea}. While MadGraph simulates the pair production of $B$, Pythia simulates its decay.  In the case of prompt signatures we add up to two additional partons in the hard-process and perform jet-parton MLM matching \cite{Alwall:2007fs}. We finally simulate the detector effects with \textsc{Delphes3}~\cite{deFavereau:2013fsa} with different configurations for ATLAS and CMS.

\subsection{Prompt signatures}

Our model resembles a simplified realisation of SUSY, where the bath particle $B$ would correspond to a slepton. Therefore, we can take advantage of LHC searches aimed at SUSY \cite{ATLAS:2019lff, CMS:2020bfa} and recast their result  to find the constraints on our model. The ATLAS and CMS analyses lead to consistent limits on SUSY models and, most importantly for our recasting, they require different (but comparable) cuts on the impact parameter relative to the primary vertex. In the following we focus on these analyses, where the same reinterpretation of prompt searches will apply to both experiments, the only difference being the exact event selection criteria.

For the ATLAS analysis in Ref.~\cite{ATLAS:2019lff} we apply all cuts for each signal region and, in particular, the ones on the impact parameter. These are $|d_0| < 5 \sigma_{d_0} \,(3 \sigma_{d_0})$ for electrons (muons), where $d_0$ is the transverse impact parameter and $\sigma_{d_0}$ the corresponding measurement error. The measurement error depends on the values of $\eta$ and $p_T$ of the tracked particle, but we can approximately take $\SI{20}{\micro\meter}$ as a reasonable average value \cite{ATLAS:2021lws,  Magliocca:2021bfg}. There is also a selection cut on the longitudinal impact parameter corresponding to $|z_0 \sin\theta|<\SI{0.5}{\milli\meter}$.  For the CMS analysis~\cite{CMS:2020bfa} we have analogous cuts on the same parameters. In particular, the search requires that $|d_0|<\SI{0.5}{\milli\meter}$ and $|z_0|<\SI{1}{\milli\meter}$. For both ATLAS and CMS we  find that the cut on the transverse impact parameter is more constraining than the one on the longitudinal impact parameter.

The fundamental parameters of our model are the LLP mass and its lifetime (or alternatively the coupling $y$). Varying the value of the coupling $y$ changes the lifetime and therefore the efficiency of the cuts on the impact parameter. However, this is the only effect of varying $y$, so we simulate our events with different mass values and a common coupling $y$. To derive the impact parameter cut efficiency for a different value of $y$ we simply rescale the measured values of $d_0$ and $z_0$ with $y^2$. We have checked that this procedure is reliable by comparing the rescaled efficiencies with the ones of events simulated at a different value of $y$.

To place limits on the parameter space, we make use of tables 8 and 9 from Ref.~\cite{ATLAS:2019lff} and table 9 from Ref.~\cite{CMS:2020bfa}. We exclude a given parameter point if the number of predicted events after cuts exceeds the quoted limit at 95\% confidence level in at least one signal region.  We note that the ATLAS analysis also employs mixed-flavour signal regions, which may be sensitive to events with two leptonically decaying tau leptons. However, we find that these signal regions are never the constraining ones since not enough events are predicted in our model anywhere in the relevant parameter space.

\subsection{Long-lived signatures}

While searches for prompt signals are sensitive to the short-distance part of the decay distribution of slightly long-lived $B$ bosons, searches for displaced leptons are more efficient for average $B$ decay lengths in the centimetre range. In the following we briefly summarise the relevant details of current ATLAS and CMS searches for displaced leptons and how we reinterpret their limits in our model.

The ATLAS collaboration has carried out a search for pairs of displaced leptons with sizeable impact parameter with a total integrated luminosity of $139$~fb$^{-1}$~\cite{ATLAS:2020wjh}. To be selected, events are required to contain two leptons with transverse impact parameter $|d_0|$ between 3~mm and 300~mm.
Events are sorted into three non-overlapping signal regions SR-$ee$, SR-$\mu\mu$ and SR-$e\mu$ according to the combination of signal lepton flavours.
Since no excess was observed, the search places a $95$~\% CL upper bound on the pair production of sleptons, both in a co-NLSP scenario, in which all three slepton flavours have the same mass, and for each single lepton flavour separately.

A similar search by CMS used $118$~fb$^{-1}$ of data in the $ee$ channel and $113$~fb$^{-1}$ in the $e\mu$ and $\mu\mu$ channels~\cite{CMS:2021kdm}. Compared to the ATLAS analysis, the CMS search is aimed at LLPs with shorter decay lengths and hence requires the transverse impact parameter $|d_0|$ of the leptons to fall between $0.1$~mm and $100$~mm.
Again, separate signal regions are defined for the electron, muon and mixed decay channel.
Like the ATLAS analysis, the CMS search sets a limit on the slepton pair production cross section, both in the co-NLSP scenario and for each slepton flavour separately.

While our model is very similar to the slepton interpretations provided by ATLAS and CMS, it differs in that our scalar $B$ only couples to right-handed leptons. The ATLAS and CMS interpretations, in contrast, assume mass-degenerate left- and right-handed sleptons. Therefore, both searches need to be reinterpreted to derive limits on our model.

Along with Ref.~\cite{ATLAS:2020wjh}, ATLAS provides 95~\% CL limits on the slepton pair production cross section as a function of the slepton mass and lifetime for the co-NLSP and each single-flavour scenario~\cite{ATLAS:2020wjh_hepdata}. These can be directly compared to the cross section for $B$ pair production in our model to exclude points in the $m_B$-$c\tau_B$ plane. 
CMS, on the other hand, provides the full cross section limit as a function of slepton mass and proper decay length only for mass-degenerate co-NLSPs~\cite{CMS:2021kdm_hepdata}. While we can proceed as for the ATLAS limit in the 3-flavour case, the single-flavour case is less straightforward. For this case, CMS only presents exclusion contours in the mass-decay-length plane but not the value of the excluded cross section in this entire parameter plane.

Since a straightforward attempt at recasting the CMS search by implementing all selection criteria does not lead to satisfactory agreement with the published limits~\cite{Araz:2021akd}, we instead infer an approximate cross section limit from the provided exclusions contours. First, we determine the cross section for the production of mass-degenerate left- and right-handed sleptons along the exclusion contour as a function of the decay length. Second, we observe that the published full cross section limit for the co-NLSP scenario has only a mild dependence on the slepton masses above approximately 300~GeV. Hence, it is a reasonable approximation to treat the single-flavour limit inferred from the exclusion contour as constant in the slepton mass for each decay length. Using this approximate limit, we can finally determine the excluded parameter region for the single-flavour case of our model as described above.

\section{Results and discussion}\label{sec:conclusions}

\begin{figure}
\centering
\includegraphics[width=0.6\textwidth]{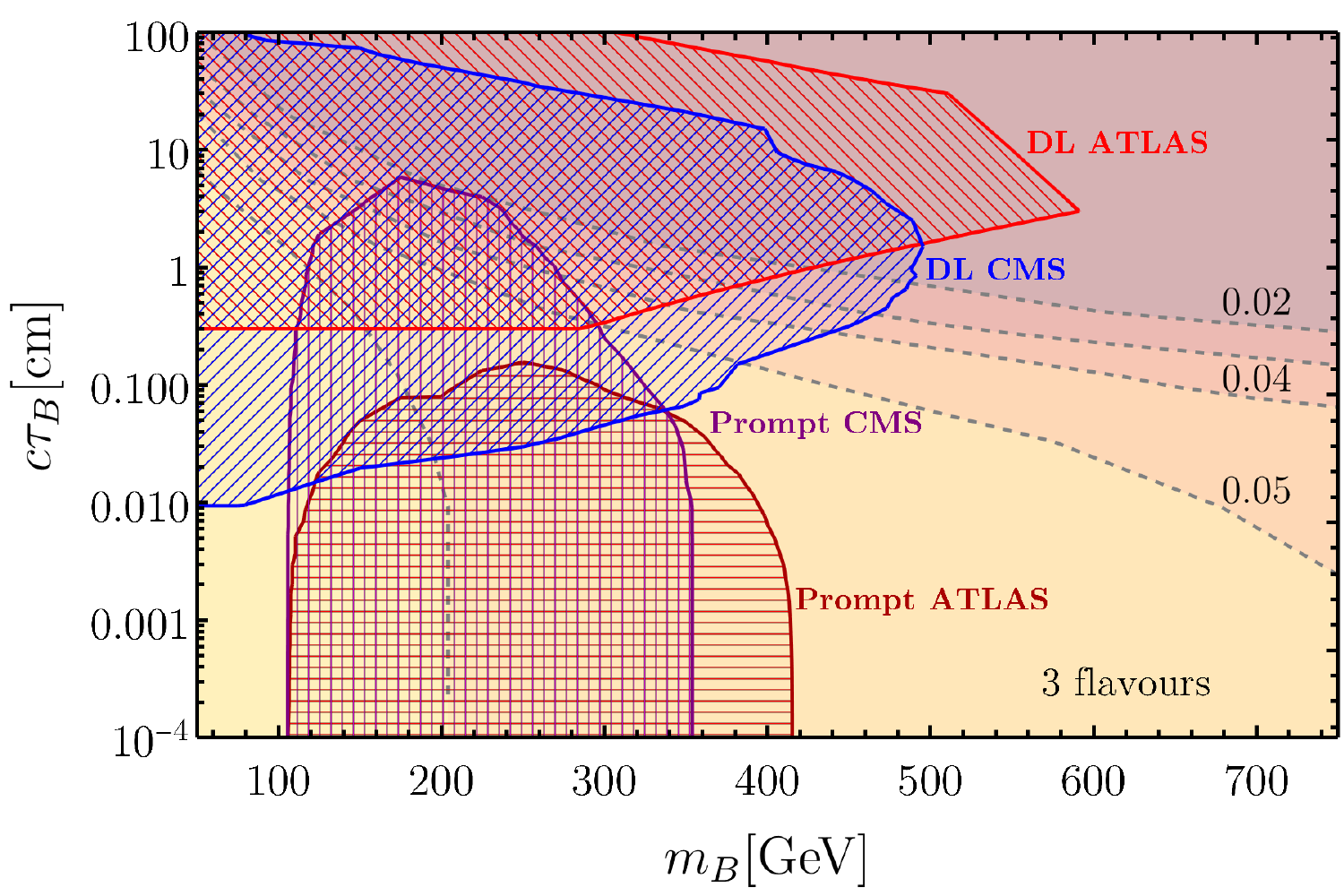}
\caption{LHC constraints on the parameter space of interest for the 3-flavour case ($y_e=y_\mu=y_\tau=y$). Note that for short lifetimes and $m_B\lesssim\SI{100}{\giga\electronvolt}$ the parameter space is at least partially excluded by LEP searches \cite{DELPHI:2003uqw}.}\label{fig:dnconstraint3f}
\end{figure}

We can now combine the predicted values of $\Delta N_\text{eff}$ from \cref{sec:freeze-in} with the LHC exclusion limits derived with the procedures outlined in \cref{sec:LHC}. The main focus of this work is on the 3-flavour case, where $y_e=y_\mu=y_\tau\equiv y$. We show the results for this scenario in \cref{fig:dnconstraint3f}. 

The first important feature of the LHC searches and one of the main results of this paper is that, once the modification of prompt limits due to macroscopic impact parameters is taken into account, the LLP and prompt exclusion limits are highly complementary. For instance, the contour for $\Delta N_\text{eff}=0.06$ can be completely excluded only when we combine the prompt and LLP constraints. This stresses the general importance of reinterpreting prompt limits for macroscopic lifetimes. By comparing CMS and ATLAS it is also noticeable how the precise value of the cuts on the impact parameters influences the sensitivity of prompt limits for longer lifetimes.

We find that the combination of searches excludes $\Delta N_\text{eff}\geq 0.055$, except for a  part of parameter space corresponding to prompt signatures and $m_B\lesssim \SI{100}{\giga\electronvolt}$, which is at least partially excluded by LEP \cite{DELPHI:2003uqw}. As a rough approximation we estimate that given the LEP centre-of-mass energy $\sqrt{s}=\SI{208}{\giga\electronvolt}$ all promptly decaying $B$ are excluded for $m_B\leq\SI{104}{\giga\electronvolt}$.

For $\Delta N_\text{eff} \lesssim 0.05$ we find that LLP searches are already more sensitive than searches for prompt decays. Given that these searches are largely free of backgrounds, substantial gains in sensitivity can be expected with increasing luminosity. In this context it will be particularly interesting to see whether ATLAS will be able to reduce the lower bound on the lepton impact parameter, which currently limits the sensitivity of its displaced lepton search in the parameter region of interest. 

We remind the reader that the results in \cref{fig:dnconstraint3f} assume that $B$ is charged only under SM hypercharge and can therefore only couple to right-handed charged leptons. If instead $B$ were an $SU(2)$ doublet and coupled to left-handed leptons, the production cross section at the LHC would increase accordingly and lead to stronger constraints. Moreover, it would also be possible to produce the second component of the $SU(2)$ doublet (analogous to sneutrinos), which would decay into $\chi$ and a SM neutrino. These decays would play no role for the LHC, since they do not produce lepton tracks, but would increase the production of DR in the early universe and thereby $\Delta N_\text{eff}$. Our calculation generalises straightforwardly to this case with no qualitative changes.

\begin{figure}
\centering
\includegraphics[width=0.49\textwidth]{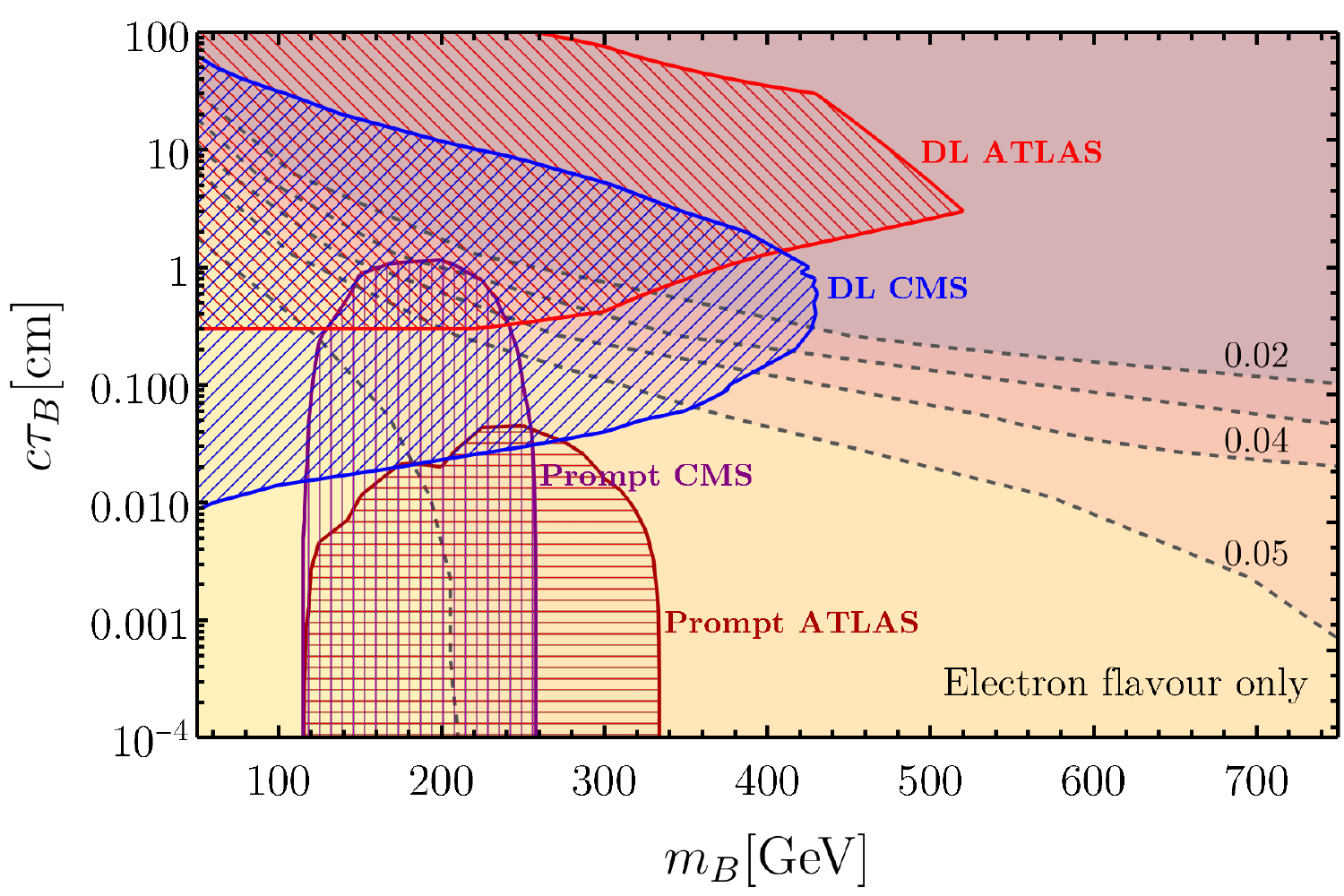}
\includegraphics[width=0.49\textwidth]{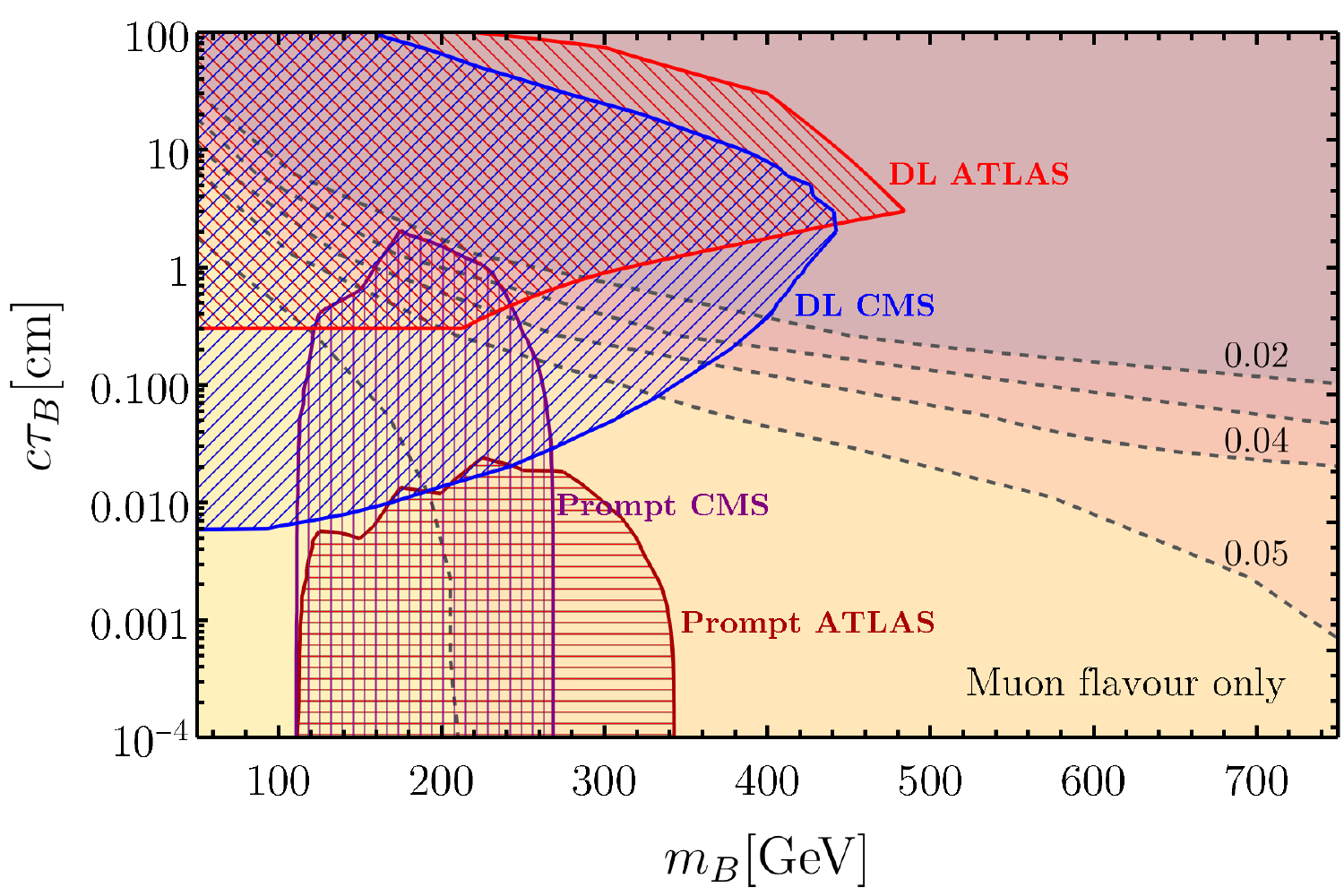}
\caption{LHC constraints on the parameter space of interest in the case of $B$ only coupling to a single flavour. On the left for coupling to electrons only  ($y_e=y$, $y_\mu=y_\tau=0$), on the right for coupling to muons only ($y_\mu=y$, $y_e=y_\tau=0$). Also in this case LEP excludes partner masses $m_B\leq \SI{104}{\giga\electronvolt}$.}\label{fig:dnconstraint1f}
\end{figure}

Another interesting generalisation of our calculation is the case where $B$ couples to only one flavour. While the analyses considered above possess good sensitivity to the case of decays into electrons or muons, they are not very sensitive to the case of tau decays. From Ref.~\cite{CMS:2021woq} we expect that even dedicated searches have only marginal sensitivity to promptly decaying $B$ coupling only to tau leptons and that this sensitivity decreases rapidly when the decays become slightly displaced. We therefore focus on the cases of couplings to right-handed electrons or muons only. The results for these scenarios are shown in \cref{fig:dnconstraint1f}.

In the single-flavour case the constraints from prompt LHC searches are weaker due to lower statistics, but also the contribution to $N_\text{eff}$ is lower, because the number of active degrees of freedom of the parent particle $B$ is reduced by a factor of 3.\footnote{By active degrees of freedom we mean the total degrees of freedom of all flavour states that couple to $\chi$ and play a role in the production of DR.  Equivalently, one can treat the degrees of freedom as constant and consider a flavour-averaged decay width, which is reduced by a factor of 3 for the single-flavour case.} Concerning the LHC limits it is worth pointing out that for the prompt ATLAS event selection there is a different requirement on muon and electron tracks, which leads to weaker bounds on the decay length of the muon partner. Furthermore, the CMS search for displaced leptons sets stronger limits in the muon channel, mainly due to lower background owing to the better impact parameter resolution for muons. In addition, the signal efficiency is slightly higher in the muon channel, especially for large $|d_0|$. In spite of these differences, \cref{fig:dnconstraint1f} shows qualitatively similar results for both cases. 

In conclusion, the interplay between cosmological observations of DR and LHC signatures of LLPs is a general result of the coincidence between the Hubble rate for temperatures around the electroweak scale and the typical vertex resolution of LHC experiments. This interplay will become particularly exciting if future CMB missions observe evidence for DR. The target sensitivity of CMB-S4 is $\sigma(N_\text{eff})=0.03$, which corresponds to our quoted $2 \sigma$ exclusion value $\Delta N_\text{eff}=0.06$, but even higher sensitivity might be reached. Looking further into the future, new experiments like CMB-HD~\cite{Sehgal:2019ewc}  aim at reaching a sensitivity of the order of $\sigma(N_\text{eff})=0.01$. With such a measurement, it may well be possible to set a \emph{lower limit} on the amount of non-thermal DR in the early universe, which would translate into a clear target for LHC experiments.

The case $\Delta N_\text{eff} \gtrsim 0.06$ can already be excluded in our model with existing LHC data. Making progress towards even smaller values of $\Delta N_\text{eff}$ will require both additional data and an improved understanding of the sensitivity of prompt searches to slightly displaced leptons. This observation justifies a more careful consideration of the event selection requirements applied by the LHC collaborations for prompt searches. Moreover, we encourage the experimental collaborations to provide the results from their prompt searches also for varying lifetimes of the parent particle, thus enabling a community-wide effort to combine the information from particle physics and cosmology.

\acknowledgments

We thank Torsten Bringmann and Jan Heisig for helpful discussions. FK and AM acknowledge funding from the Deutsche Forschungsgemeinschaft (DFG) through the Collaborative Research Center TRR 257 ``Particle Physics Phenomenology after the Higgs Discovery'' under Grant 396021762 -- TRR 257 and the Emmy Noether Grant No. KA 4662/1-1. ML is funded by the Alexander von Humboldt Foundation. This manuscript has been authored by Fermi Research Alliance, LLC under Contract No.~DE-AC02-07CH11359 with the U.S.\ Department of Energy, Office of Science, Office of High Energy Physics.

\bibliographystyle{JHEP_improved}
\bibliography{bibliography}

\end{document}